\documentclass{article}
\usepackage{hiph-art}
\usepackage{graphicx} 
\usepackage{amsmath,amssymb,epsfig,bm}
\recrevdate{25 November 2005}                                              
\def\rightmark{Transport coefficients from string theory}
\def\leftmark{A.O.~Starinets}

\newcommand{\<}{\langle}
\renewcommand{\>}{\rangle}

\newcommand{\q}{\bm{q}}
\newcommand{\x}{\bm{x}}

\def\wf{Z} 

\title{Transport coefficients of strongly coupled gauge 
theories: insights from string theory} 
\authors{ 
{Andrei O. Starinets%
\index{One, A.} 
\index{Two, A.} 
}\\[2.812mm]
{\normalsize
\hspace*{-8pt}$\,\;$ Perimeter Institute for Theoretical Physics,\\ 
31 Caroline Street North, Waterloo, Ontario, N2L 2Y5, Canada\\
{\it e-mail: starina@perimeterinstitute.ca}\\[0.2ex] 
}}
 
\abstract{The transport properties of certain 
strongly coupled thermal gauge theories can be determined from their 
effective description in terms of gravity or superstring theory duals.
Here we provide a short summary of the results for the shear and bulk viscosity, 
charge diffusion constant, and the speed of sound in 
supersymmetric strongly interacting plasmas. We also outline a general
 algorithm for computing transport coefficients in any gravity dual. 
The algorithm relates the transport coefficients to the coefficients
 in the quasinormal spectrum of five-dimensional 
black holes in asymptotically anti de Sitter space.}
\keyword{AdS/CFT correspondence, thermal field theory}

\PACS{11.25.Hf, 11.10.Wx }
 
\makeindex
\begin{document}
 
\maketitle

\section{Introduction}\label{intro}
The gauge theory - gravity or, more generally, gauge theory - string theory
 duality  \cite{Maldacena:1997re}, \cite{Gubser:1998bc}, \cite{Witten:1998qj} (for reviews, see
\cite{Aharony:1999ti}, \cite{Klebanov:1999ku}, \cite{Klebanov:2000me})
 has been developed over the last eight years 
as a novel rigorous approach for understanding non-perturbative aspects of gauge theories. Together with lattice  gauge theory, it provides a variety of useful insights into the dynamics of strongly coupled quantum systems. The major advantage of the gauge/gravity duality 
over the lattice methods is that it allows one to obtain precise analytical results in the regime of strong coupling. Its major disadvantage is that
 only a limited number of gauge theories, namely the ones for which gravity- or string theory duals are known, can be treated in this approach. Unfortunately, this class of theories does not yet include QCD,  
 although there seems to be a consensus that a string theory dual to QCD should 
exist (see e.g.~\cite{polchinski_talk}, \cite{Klebanov:2005mh}). Even in the absence of a dual string theory description of QCD, results for other gauge theories and in particular thermal gauge theories,
 are of interest: first, they have a theoretical significance of their own; second, 
they may uncover universal properties of the strong coupling dynamics shared by a large class of (if not all) gauge theories.

It is important to emphasize that the duality is a {\it conjecture} rather than 
an established mathematical or experimental fact. To date, it survived 
numerous theoretical tests and is generally believed 
 to be true, although it is unlikely that 
a formal mathematical proof will ever be given.  
It is thus important to seek independent verifications of the
 results provided by the correspondence; in particular, a direct comparison with 
lattice gauge theory results would be highly desirable.

The original example of the gauge-gravity duality conjecture 
(also known as the AdS/CFT correspondence) was proposed 
in 1997 by J.~Maldacena \cite{Maldacena:1997re}.
It is a statement that a certain (type IIB) string theory on $AdS_5\times S^5$ (five-dimensional anti de Sitter space times a five-dimensional sphere) is {\it equivalent}
 to ${\cal N}=4$ supersymmetric $SU(N_c)$ Yang-Mills (SYM) theory in four dimensions.
The ``extra'' six dimensions of  $AdS_5\times S^5$ are interpreted as internal parameters of the Yang-Mills theory: the fifth dimension of the  anti de Sitter space
is related to the energy scale of the gauge theory, and the rotational group $SO(6)$ 
of the sphere is precisely the $R$-symmetry group of the ${\cal N}=4$ SYM. 
 The ${\cal N}=4$ SYM is a superconformal 
theory with a vanishing beta-function. Thus, unlike in QCD,  the coupling does not run 
and can be set to any desired value. The theory does not have any internal
 scale (such as $\Lambda_{QCD}$) and is not confining.

As stated, the conjectured equivalence is supposed to be valid for 
any range of the parameters of the gauge theory (i.e. for any $N_c$ and any coupling).
The limit of (infinitely) large $N_c$ and (infinitely) 
large 't Hooft coupling $g_{YM}^2 N_c$ in
 the gauge theory corresponds (on the string theory side of the duality)  
to a classical (super)gravity approximation to the full string theory.
Most calculations in AdS/CFT are done in this limit. Corrections in 
powers of inverse 't Hooft coupling can be computed by considering stringy corrections 
to the classical anti de Sitter geometry. Corrections in inverse powers of $N_c$ 
correspond to quantum gravity corrections. Finding these corrections is a difficult problem. 

Gravity duals of non-conformal theories can also be constructed 
(see e.g.~\cite{Girardello:1998pd}, \cite{Polchinski:2000uf}). 
These non-conformal theories are usually the mass deformations of the original 
 ${\cal N}=4$ SYM theory (a ``mass deformation'' here means that 
the IR-relevant operators such as mass terms are added to the 
 ${\cal N}=4$ SYM Lagrangian thus introducing a scale $M$ into the theory). 
In the UV, i.e. for energies $E\gg M$, such theories reduce to  ${\cal N}=4$ SYM.
In the IR, they become  ${\cal N}=2$,  ${\cal N}=1$ SYM or a
 non-supersymmetric gauge theory - depending on the  particular type of a mass 
deformation.
To emphasize the difference in the behavior of
 such theories above and below the mass deformation scale $M$, they are denoted 
${\cal N}=2^*$,  ${\cal N}=1^*$. These theories can exhibit confinement, chiral symmetry breaking and other interesting features - all of them identified and accessible to
 study through their effective gravity 
duals\footnote{Quantitative study of these gravity duals may face 
formidable technical difficulties.}. By shifting the mass deformation scale to the UV,
one can - at least in principle - make these theories arbitrarily close to 
the ``realistic'' gauge theories (e.g.  ${\cal N}=1$ SYM).
However, as the coupling becomes small in the UV, the dual classical 
gravity description breaks down and must be replaced by a 
full string theory description which is not understood at the moment.

Thermal properties of the gauge theories can be studied by including black holes in the 
(asymptotically) AdS background
 \cite{Maldacena:1997re}, \cite{Witten:1998zw}. The Hawking temperature and the Bekenstein-Hawking entropy of the gravity background are  
identified, respectively,  with the temperature and entropy 
of the gauge theory in thermal equilibrium.
One of the earliest results in thermal AdS/CFT was the calculation of the entropy of 
${\cal N}=4$ SYM in the regime of large $N_c$ and large 't Hooft coupling 
\cite{Gubser:1996de}.

In thermal gauge theories, 
of particular interest is the regime described by hydrodynamics.
This near-equilibrium regime is completely characterized 
by  transport coefficients (e.g. shear and bulk viscosities)
whose values are determined by the  
underlying microscopic theory. In practice they are hard to 
compute from ``first principles'', even in perturbation theory. (For example, 
no perturbative calculation of the bulk viscosity in a gauge theory 
seems to be 
 available at the moment.) Lattice approaches  to computing transport
 coefficients (see e.g.~\cite{Karsch:1986cq}, 
\cite{Sakai:2005fa}) rely on indirect methods and to the best of our knowledge 
cannot yet provide unambiguous quantitative results.
 Thus the AdS/CFT remains the only 
source of theoretical information about transport properties of thermal 
gauge theories in the nonperturbative regime, although the gauge theories 
in question do not include QCD.

In order to study the near-equilibrium processes needed for the calculation
 of transport properties in AdS/CFT, one considers perturbations of the black hole background.
These perturbations correspond to deviations from the thermal equilibrium
 in the dual gauge theory.

In this short review we provide a summary of the  current state of affairs in 
computing the transport properties from gravity duals.
More details can be found in the original articles 
\cite{Policastro:2001yc} --- \cite{Benincasa:2005iv}.

\section{Methods for computing transport coefficients from dual gravity}

Gauge/gravity duality allows one to compute the correlation functions 
of gauge-invariant operators from gravity. To deal with real time processes 
in the gauge theory one needs the Lorentzian rather than Euclidean 
formulation of the correspondence. A concrete Lorentzian recipe 
for the two-point functions was given in \cite{Son:2002sd}
and generalized to higher point functions in 
\cite{Herzog:2002pc}.

\subsection{Viscosity from the  Green-Kubo formula}
Once the retarded correlation functions are known, transport coefficients 
can be obtained from Green-Kubo formulas. For example, for the shear viscosity
we have 
\begin{equation}\label{kubo}
\eta = \lim_{\omega \rightarrow 0} \frac{1}{ 2\omega} 
 \int\!dt\,d\x\, e^{i\omega t}\,
\langle [ T_{xy}(x),\,  T_{xy}(0)]\rangle = 
- \lim_{\omega \rightarrow 0} {\mbox{Im}\; 
\frac{G^R_{xy,xy}(\omega,0)}{ \omega}}\,,
\end{equation}
where  the
retarded Green's function for the components of the stress-energy tensor
is defined as
\begin{equation}
  G^R_{\mu\nu,\lambda\rho} (\omega, \q)
  = -i\!\int\!d^4x\,e^{-iq\cdot x}\,
  \theta(t) \< [T_{\mu\nu}(x),\, T_{\lambda\rho}(0)] \>\,.
\end{equation}
Thus finding the shear viscosity is equivalent to  
computing the zero-frequency limit of the imaginary part of the 
retarded correlator $G_{xy,xy}^R$.

Such a computation can be carried out explicitly in the simplest example of 
thermal  ${\cal N}=4$ $SU(N_c)$ SYM in the limit of infinite $N_c$ and infinite 't Hooft coupling $g^2_{YM} N_c$. An equilibrium thermal state of the theory at temperature $T$ 
is described by a dual five-dimensional AdS-Schwarzschild black hole 
(more precisely, by a black hole with a
translationally invariant horizon) whose metric is given by 
\begin{equation}\label{near_horizon_metric}
ds^2_{5} = 
  \frac{(\pi T R)^2}u
\left( -f(u) dt^2 + dx^2 + dy^2 +dz^2 \right) 
 +\frac{R^2}{ 4 u^2 f(u)} du^2\,,
\end{equation}
where $f(u)=1-u^2$, $T$ is the Hawking temperature, $R$ is the AdS radius.
In Eq.~(\ref{near_horizon_metric}), the horizon corresponds to $u=1$,
the spatial infinity to $u=0$. One may think of the four-dimensional gauge
 theory as of a theory defined on the ``boundary'' 
at $u\rightarrow 0$ of the spacetime
(\ref{near_horizon_metric}) with the standard Minkowski coordinates $t,x,y,z$.

According to the Lorentzian version of the AdS/CFT, the retarded correlator
$G_{xy,xy}^R$ is completely known (in the limit 
 $g^2_{YM} N_c\rightarrow \infty$, $N_c\rightarrow \infty$) if the solution 
to the 
equation for a minimally coupled massless scalar propagating in the background 
(\ref{near_horizon_metric}) is known. The second-order differential equation 
should of course be supplemented by the appropriate boundary conditions, as 
discussed in \cite{Son:2002sd}. Different boundary conditions at the (future) 
horizon correspond to different types of Green's 
functions (retarded, advanced, Feynman etc.) in the field theory.
 The
(retarded) two-point function 
of the stress-energy tensor (in Fourier space) in 
the hydrodynamic approximation reads 
\begin{equation} 
  G_{xy,xy}(\omega,\q) =
  - \frac{i \pi N_c^2 \omega  T^3 }{ 8 }  \,.
\end{equation}
Using the Kubo formula (\ref{kubo}) for the shear viscosity one obtains 
\cite{Policastro:2001yc}, \cite{Policastro:2002se}
\begin{equation}
\eta =  \frac{\pi}8 N_c^2 T^3\,.
\label{visc_0}
\end{equation}
Following 't Hooft's philosophy of the large-$N_c$ 
expansion \cite{'tHooft:1973jz}, the generic expression for the shear viscosity in conformal gauge 
theory  can be written as 
\begin{equation}
\eta = T^3 \, \sum_{k=0}^{\infty} \, N_c^{2-2k}\, f_k (g^2_{YM} N_c)\,,
\end{equation}
where the dependence on temperature follows from the dimensional analysis 
and the fact that for a
 thermal conformal theory the only scale is the temperature itself. 
The result  (\ref{visc_0}) means that 
\begin{equation}
\lim_{g^2_{YM} N_c \rightarrow \infty} f_0
=\pi/8\, .
\end{equation}
Taking into account classical stringy corrections to the metric 
(\ref{near_horizon_metric})  (i.e. the corrections arising from the
 fact that strings are one-dimensional rather than point-like objects)
one can compute the correction to the result (\ref{visc_0}) \cite{Buchel:2004di}:
\begin{equation}
\eta =  \frac{\pi}8 N^2 T^3 \left( 1 + 
\frac{75}{4} \zeta(3) \, (2 g^2_{YM} N_c)^{-3/2}+\cdots \right)\,,
\label{visc_corr}
\end{equation}
where $\zeta(3)\approx 1.2020569$ is Ap\'{e}ry's constant.
Combining Eq.~(\ref{visc_corr}) with the result for the strong coupling 
limit of the volume entropy density \cite{Gubser:1998nz}
\begin{equation}
s = S/V_3 = \frac{2\pi^2}{3}\, N_c^2 T^3 \left( \frac{3}{4} +  \frac{45}{32}
 (2 g^2_{YM} N_c)^{-3/2}+\cdots \right)\,,
\end{equation}
we find for  $g^2_{YM} N_c \gg 1$
\begin{equation}
\frac{\eta}{s} = \frac{1}{4\pi} \left( 1 + 
\frac{135}{8} \, \zeta(3) \, (2 g^2_{YM} N_c)^{-3/2}+\cdots \right)\,.
\label{visc_corr_a}
\end{equation}
On the other hand, 
 perturbative field theory calculations at weak coupling $g\ll1$ 
give (see e.g. \cite{Arnold:2000dr}
and references therein)
\begin{equation}
\frac{\eta}{s} \sim \frac{1}{g^4 \log{1/g^2}}\,.
\end{equation}
The dependence of $\eta/s$ on 't Hooft 
coupling in  ${\cal N}=4$ $SU(N_c)$ SYM (at infinite $N_c$) is shown schematically 
in Fig. 1.

\begin{figure}[tb]
\centerline{\includegraphics[width=11cm,angle=0]{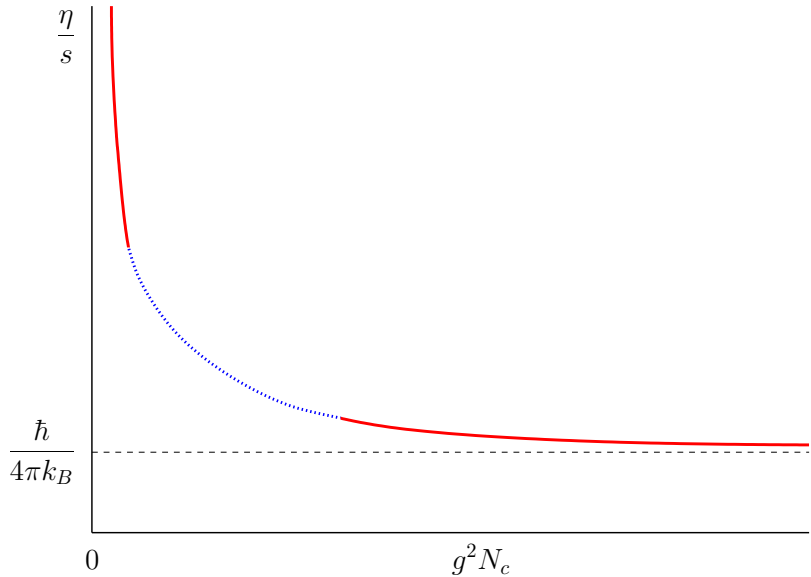}}
\caption{\label{fig1} The dependence 
of the ratio $\eta/s$ on the 't Hooft coupling
$g^2N_c$ in ${\cal N}=4$ supersymmetric Yang--Mills theory.  The ratio
diverges in the limit $g^2 N_c\to0$ and approaches 
$\hbar/{4 \pi k_B}$
from above as $g^2N_c\to\infty$.  
The ratio is unknown in the regime
of intermediate 
't~Hooft coupling $g^2 N_c\sim1$, but is likely to be a 
monotonic function of $g^2 N_c$. (Adapted from Ref.~\cite{Kovtun:2004de}.)}
\end{figure}

\subsection{Transport coefficients from hydrodynamic 
poles of the retarded correlators}

Hydrodynamics predicts that the retarded two-point correlation functions
involving densities of conserved quantities must have singularities whose 
dispersion relation satisfy $\omega(q)\rightarrow 0$ as $q\rightarrow 0$,
where $q$ is the magnitude of the spatial momentum.
For example, correlators of the components of the stress-energy 
tensor associated with 
the transverse fluctuation of the momentum density should exhibit a diffusion pole in the hydrodynamic regime,
\begin{equation}
G_{tx,xz}(\omega,\q) \sim 
- \frac{ N_c^2\pi T^3 \omega q}{  8 (i\omega - {\cal D} q^2)}\,,
\end{equation}
where the corresponding dispersion relation
\begin{equation}
\omega (q) = - i\, {\cal D}\, q^2 +  O(q^3)\,
\label{aa1}
\end{equation}
involves the diffusion constant
 ${\cal D} = \eta/(\varepsilon +P)$
which depends on
 the shear viscosity as well as on 
the energy density $\varepsilon$ and the pressure $P$. (For homogeneous systems with zero chemical potential $\varepsilon + P = s T$.)

Similarly, correlators of the diagonal components of the stress-energy 
tensor corresponding to the propagation of the sound waves 
in a Yang-Mills plasma have a hydrodynamic pole with the dispersion relation
\begin{equation}
\omega (q) = v_s\, q - i \frac{\Gamma}{2} \, q^2 + O(q^3)\,,
\label{aa2}
\end{equation}
where $v_s = (\partial P/\partial \varepsilon)^{1/2}$
is the speed of sound. The sound attenuation constant 
$\Gamma$ depends on shear and bulk viscosities $\eta$ and $\zeta$:
$$
\Gamma = \frac{1}{\varepsilon +P} \left( \zeta + \frac{4}{3} \eta \right)\,.
$$
Correlators $G_{\mu\nu,\sigma\rho}$ 
(as well as the correlators of the $R$-currents) in  ${\cal N}=4$  SYM 
(at infinite $N_c$ and infinite 't Hooft coupling)  
can be computed analytically in the hydrodynamic approximation by 
solving the corresponding system of 
differential equations obeyed by the dual gravitational fluctuations.
 The functional form of the correlators coincides with the one 
expected from hydrodynamics. From the dispersion relations  
such as the ones given by Eqs.~(\ref{aa1}), (\ref{aa2}) we identify the 
transport coefficients of  ${\cal N}=4$ SYM in the above mentioned limit 
\cite{Policastro:2002se}, \cite{Policastro:2002tn} :
\begin{equation}
\eta = \frac{\pi}{8}\,  N_c^2 T^3 \,, \; \qquad \; \zeta =0\,, \;
 \qquad \; v_s = \frac{1}{\sqrt{3}}\,, \qquad \; D_R = \frac{1}{2\pi T}\,,
\end{equation}
where $D_R$ is the R-charge diffusion constant.

The procedure of extracting the transport coefficients form the poles of the 
retarded correlators can be generalized to include any gravity
 dual \cite{Kovtun:2005ev}.
At finite temperature, a generic stress-energy tensor two-point function depends  (up to an index structure) on five scalar functions. 
On the gravity side,
one can identify five gauge-invariant combinations of the metric perturbations
that obey a coupled system of differential equations.
The dispersion relations (\ref{aa1}), (\ref{aa2}) appear as the 
lowest eigenfrequencies of this system 
(the so called quasinormal frequencies). Finding the  quasinormal spectrum
is a well posed (although often difficult) problem  in mathematical physics.
Examples of application of this general formalism to non-conformal theories 
can be found in \cite{Parnachev:2005hh}, \cite{Benincasa:2005iv}.
In particular, in \cite{Benincasa:2005iv} 
for the strongly coupled ${\cal N}=2^*$ SYM it was found that while the 
ratio $\eta/s$ remains equal to $1/4\pi$, 
the bulk viscosity is non-zero and is proportional
to the deviation of the speed of sound squared from its value in conformal theory,
\begin{equation}
\frac{\zeta}{\eta} \simeq \kappa \left( v_s^2 - \frac{1}{3}\right)\,,
\end{equation}
where $\kappa$ is a coefficient of order one. Transport properties of 
strongly coupled gauge theories with a spontaneously generated mass scale
 (the so called cascading gauge theories) were recently studied in 
\cite{Buchel:2005cv}.

\subsection{The ``membrane paradigm'' approach}
For a gravity dual with a metric of the form 
\begin{equation}\label{full_membrane}
  ds^2 = G_{00}(r)\,dt^2 + G_{rr}(r)\,dr^2 + 
         G_{xx}(r) \sum_{i=1}^{p} (dx^i)^2 
         + \, \wf(r) \, K_{mn}(y) \, dy^m dy^n\,,
\end{equation}
where the components $G_{00}(r)$, $G_{rr}(r)$, $G_{xx}(r)$,  
and the ``warping factor'' $\wf(r)$ 
depend only on the radial coordinate $r$, one can derive a 
generic formula
for a diffusion coefficient \cite{Kovtun:2003wp}
\begin{equation}
  D = 
  \frac{\sqrt{-G(r_0)} \, \wf(r_0)}{G_{xx}(r_0) 
        \sqrt{-G_{00}(r_0) \, G_{rr}(r_0)}}
        \int\limits_{r_0}^\infty\!dr\, 
     \frac{-G_{00}(r) \, G_{rr}(r)}{\sqrt{-G(r)} \, \wf(r)}\,
\label{main_D}
\end{equation}
 in terms of the metric components computed 
at the horizon $r=r_0$. Similarly, for the shear viscosity one has
\begin{equation}\label{eta}
  \eta = s\,T \frac{\sqrt{-G(r_0)}}{\sqrt{-G_{00}(r_0) \, G_{rr}(r_0)}}
  \int\limits_{r_0}^\infty\!dr\,
  \frac{-G_{00}(r) \, G_{rr}(r)}{G_{xx}(r) \sqrt{-G(r)}} \,.
\end{equation}
These formulas were derived in the old, 
``pre-AdS/CFT'' framework. However, it can be shown \cite{starinets_appear}
 that the same results
arise from the computation of the lowest quasinormal frequency in the background (\ref{full_membrane}).

\section{Universality of shear viscosity in (super)gravity approximation}

Calculations based on the Green-Kubo formula \cite{Kovtun:2004de},
 the shear mode pole in the stress-energy tensor correlators 
\cite{Buchel:2004qq} or
 the generic formula (\ref{eta}) \cite{Buchel:2003tz}
all reveal a rather surprising result that  the 
ratio of the shear viscosity to the volume entropy density is universal 
and equal to $1/4\pi$ for {\it any} gauge theory in the regime of coupling 
and other parameters described by a {\it gravity} dual.
Note that this statement says nothing about the behavior of
 the viscosity beyond the dual gravity approximation. 
(For example, for ${\cal N}=4$ SYM 
one expects corrections to the ratio to appear 
for finite $N_c$ and $g_{YN}^2 N_c$, as illustrated by
 Eq.~(\ref{visc_corr_a}). Such corrections are not expected to be universal.) 

\begin{figure}[tb]
\vspace*{1cm}
\centerline{\includegraphics[width=11cm,angle=0]{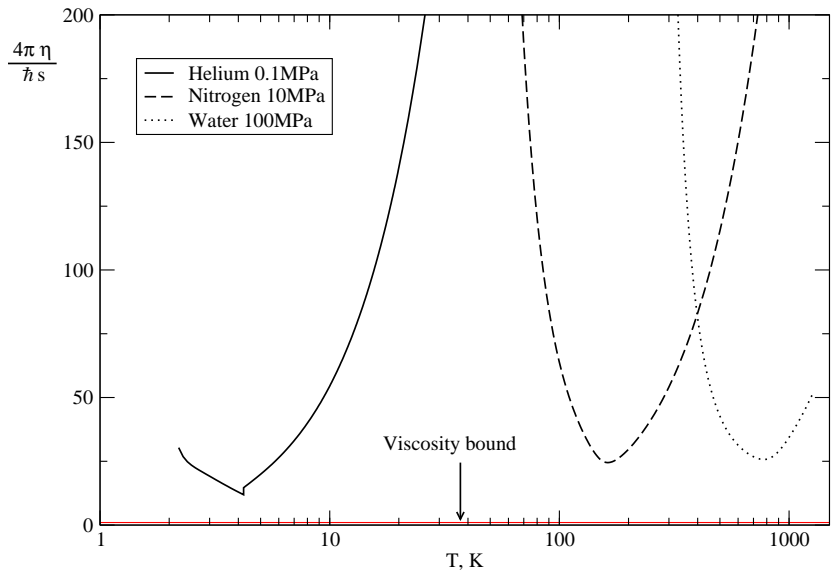}
}
\caption{\label{fig2}The viscosity-entropy ratio for some common substances:
helium, nitrogen and water.  The ratio is always substantially larger
than its value in theories with gravity duals, represented by the
horizontal line marked ``viscosity bound''. 
(Adapted from Ref.~\cite{Kovtun:2004de}.)}
\end{figure}

The regime described by gravity duals 
is normally associated with an infinitely strong coupling. 
At the same time, at weak coupling the ratio $\eta/s$ is typically very large.
Restoring fundamental  constants, we see that 
$\eta/s=\hbar/4\pi k_B \approx 6.08 \cdot  10^{-13} K \cdot s$ is 
independent of the speed of light or the gravitational constant. 
One may put forward a {\it speculation}
that perhaps the 
 limit $1/4\pi$ serves as a universal lower bound on the value of 
$\eta/s$ \cite{Kovtun:2003wp}. Arguments in favor of this speculation can be found in 
\cite{Kovtun:2003wp},
\cite{Kovtun:2004de}.
Experimental data for $\eta/s$ for some common liquids are shown in Fig.~2.
\section{Conclusions}
We have outlined a general approach for computing the transport coefficients 
of strongly coupled gauge theories from gravity. 
The ratio of shear viscosity to entropy density 
in the regime described by a gravity dual is 
universal and equal to $1/4\pi$ which is significantly lower
 than the corresponding
perturbative result.
Hydrodynamic models  used to describe the
 elliptic flows observed 
 at RHIC seem to suggest \cite{Teaney:2003pb}, \cite{Shuryak:2003xe} 
that the ratio  $\eta/s$ in the 
non-perturbative regime of QCD should be small and relatively 
 close to  $1/4\pi$. 
Whether or not this intriguing observation is indeed related to 
the universality result for gravity duals remains to be seen.

\section*{Acknowledgments}
The reviewed results were obtained in collaboration with D.~T.~Son, 
G.~Policastro, A.~N\'{u}\~{n}ez,
 P.~Kovtun, A.~Buchel, J.~Liu, A.~Parnachev and 
P.~Benincasa.
I would like to thank the organizers and especially 
A.~Rebhan for the invitation to speak at the ``Quark-Gluon Plasma
Thermalization'' workshop at the Technische Universit\"{a}t Wien
and for their warm hospitality. I would also like to thank the 
participants of the workshop and particularly 
L.G.~Yaffe for very useful discussions.

\vfill\eject
\end{document}